\documentclass[aps,shownopacs,superscriptaddress,nofootinbib,preprint,11pt]{revtex4}
\usepackage{hyperref}
\usepackage{amsmath}
 \usepackage{multirow}
\usepackage{array}
\newcolumntype{L}[1]{>{\raggedright\let\newline\\\arraybacksslash\hspace{0pt}}m{#1}}
\newcolumntype{C}[1]{>{\centering\let\newline\\\arraybackslash\hspace{0pt}}m{#1}}
\newcolumntype{R}[1]{>{\raggedleft\let\newline\\\arraybackslash\hspace{0pt}}m{#1}}
\usepackage{float}
\usepackage{graphicx}
\usepackage{epsfig}
\usepackage{psfrag}
\usepackage{color}
\usepackage{slashed}

\usepackage{amsfonts}
\usepackage{amssymb}
\usepackage{tikz}
\usepackage{tikz}
\usetikzlibrary{positioning,arrows}
\usetikzlibrary{decorations.pathmorphing}
\usetikzlibrary{decorations.markings}

\newcommand*{\be}{\begin{equation}}
\newcommand*{\ee}{\end{equation}}
\newcommand*{\bea}{\begin{eqnarray}}
\newcommand*{\eea}{\end{eqnarray}}

\newcommand{\comment}[1]{}

\newcommand{\cref}[1]{Chapter~\ref{c.#1}}

\def\beq{\begin{equation}}
\def\eeq{\end{equation}}
\def\bea{\begin{eqnarray}}
\def\eea{\end{eqnarray}}
\def\ba{\begin{array}}
\def\ea{\end{array}}
\def\bi{\begin{itemize}}
\def\ei{\end{itemize}}
\def\be{\begin{enumerate}}
\def\ee{\end{enumerate}}
\def\bc{\begin{center}}
\def\ec{\end{center}}
\def\bt{\begin{table}}
\def\et{\end{table}}
\def\btb{\begin{tabular}}
\def\etb{\end{tabular}}

\def\simgt{\stackrel{>}{{}_\sim}}

\def\lsim{\raise0.3ex\hbox{$\;<$\kern-0.75em\raise-1.1ex\hbox{$\sim\;$}}}
\def\gsim{\raise0.3ex\hbox{$\;>$\kern-0.75em\raise-1.1ex\hbox{$\sim\;$}}}
	
 \preprint{ DAMTP-2017-28,\ 
 	TIFR/TH/17-29
 }
\begin{document}

\title{Dissecting Multi-Photon Resonances at the Large Hadron Collider}

\author{B.C. Allanach}
\email{b.c.allanach@damtp.cam.ac.uk}
\address{Department of Applied Mathematics and Theoretical Physics, Centre for Mathematical Sciences,\\
	University of Cambridge, Wilberforce Road, Cambridge, CB3 0WA, United Kingdom}

\author{D. Bhatia}
\email{disha@theory.tifr.res.in}
\address{Department of Theoretical Physics, Tata Institute of Fundamental Research, Homi Bhabha Road, Colaba, Mumbai 400 005, India}

\author{Abhishek M. Iyer}
\email{abhishek@theory.tifr.res.in}
\address{Department of Theoretical Physics, Tata Institute of Fundamental Research, Homi Bhabha Road, Colaba, Mumbai 400 005, India}

\begin{abstract}
We examine the phenomenology of the production, at the 13 TeV Large Hadron
Collider (LHC),  of a heavy resonance $X$, 
which decays via other new on-shell particles $n$ into multi- (i.e.\ three
or more) photon final states.  
In the limit that $n$ has a much smaller mass than $X$, 
the multi-photon final state may dominantly appear as a two photon
final state because the $\gamma$s from the $n$ decay are highly
collinear and remain unresolved.
We discuss how to discriminate this scenario
from $X \rightarrow \gamma \gamma$: rather than discarding non-isolated
photons, it is better instead to relax the isolation criterion and instead
form photon 
jet substructure variables. 
The spins of $X$ and $n$ leave their imprint upon
the distribution of pseudorapidity gap $\Delta \eta$ between the apparent two
photon 
states. Depending on the total integrated luminosity,  this can be used
in many cases
to claim discrimination between the possible spin choices of $X$ and $n$,
although 
the case where $X$ and $n$ are both scalar particles
cannot be discriminated from the direct $X \rightarrow \gamma \gamma$ decay in
this manner. 
Information on the mass
of $n$ can be gained by considering the mass of each photon jet. 
\end{abstract}

\maketitle

\section{Introduction}
\label{Introduction}
\begin{figure}
	\begin{center}
		\includegraphics[width=\textwidth]{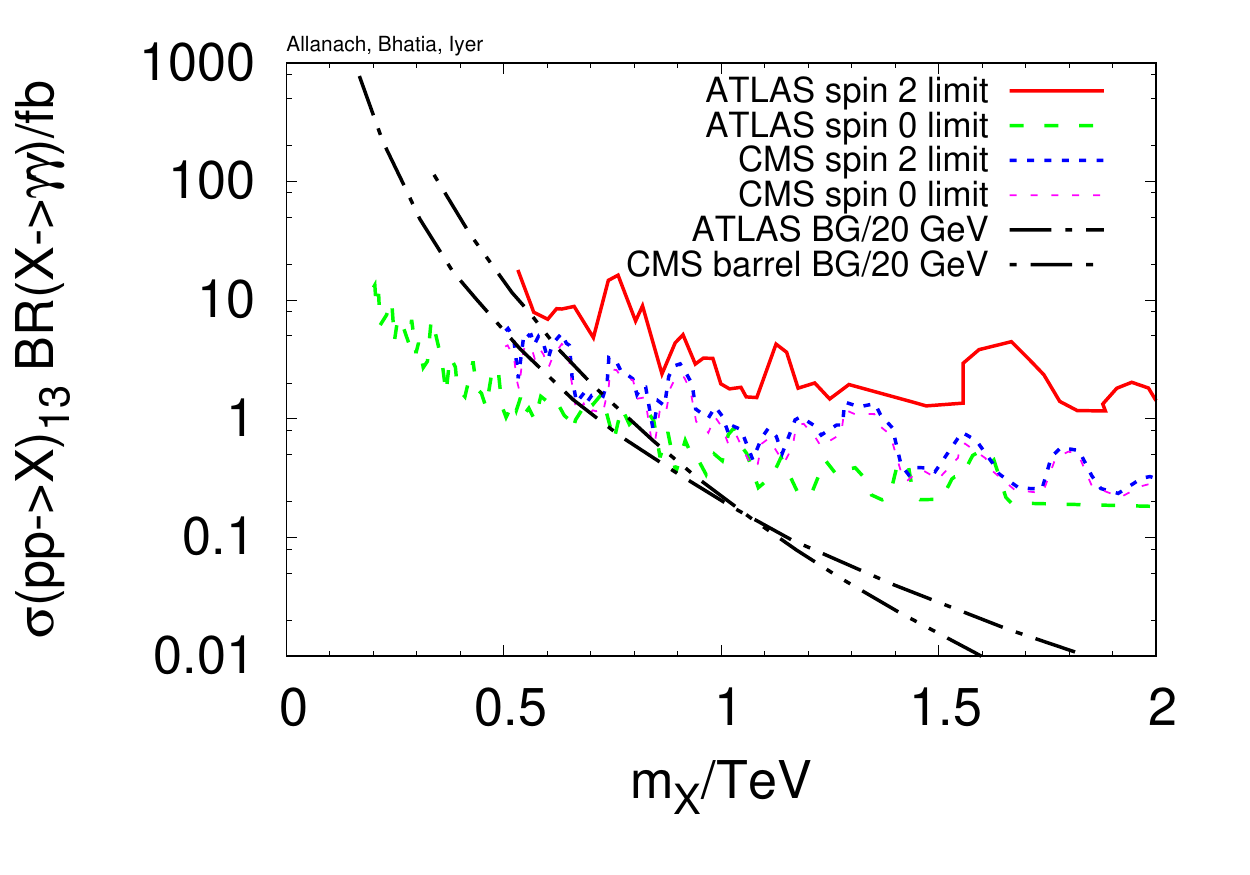}
	\end{center}
	\caption{Upper limits on 13 TeV LHC di-photon resonance production and
          fitted           backgrounds 
          for the di-photon invariant mass spectrum.
In the curves marked ``limit'', we display the upper 95$\%$ confidence level
limit on the 
cross-section times branching
		ratio of a narrow resonance that decays into a two photon final state. The ATLAS spin~0 limits were obtained from 15.4
		fb$^{-1}$ of 
		integrated luminosity~\cite{ATLAS:2016eeo}, the
		ATLAS spin~2 limits came from 
		3.2 fb$^{-1}$~\cite{Aaboud:2016tru} under the assumption of 
		a Randall Sundrum graviton~\cite{Randall:1999ee}, whereas the
                CMS limits come 
		from a 
		combination of 19.7 fb$^{-1}$ of 8 TeV collisions and 15.2
                fb$^{-1}$ of 13 TeV 
		collisions~\cite{Khachatryan:2016yec}. The curves labelled ``BG'' show central
                values of fitted di-photon mass spectra for 13 TeV 
           LHC collisions in a  3.2 fb$^{-1}$ ATLAS
           analysis~\cite{Aaboud:2016tru} and for a 
           12.9 fb$^{-1}$ CMS analysis~\ cite{Khachatryan:2016yec} where both
           photons end up in the barrel. The expected background (`BG') in
           each case is shown for a bin of width 20 GeV.}
	\label{fig:sigLim}
\end{figure}

The Standard Model (SM) of particle physics has been extensively tested to a great
degree of accuracy. 
 The discovery of a particle whose properties are so far
consistent with those predicted for the SM Higgs boson have further fuelled
the searches for Beyond the Standard Model (BSM) physics.
 The  typical signatures
employed in the search for these 
new physics scenarios involve different combinations of hard
isolated photons,
hard jets, hard isolated leptons and large missing transverse momentum. The
presence of isolated leptons 
and isolated photons in a given final state is useful in significantly
depleting SM  backgrounds.
The discovery of the Higgs boson in the di-photon
channel~\cite{Aad:2012tfa,Chatrchyan:2012xdj} has lead to an 
increased interest in the $\gamma \gamma$ final state. A hunt
for a putative heavy resonance $X$ enjoys enhanced sensitivity 
because SM backgrounds reduce quickly at larger di-photon 
invariant masses $m_{\gamma\gamma}$. Fits to the $m_{\gamma \gamma}$ distribution
are obtained  by both the
ATLAS and CMS by assuming simple functional forms. The central values of the
fitted forms for 13 TeV LHC collisions are shown in 
Fig.~\ref{fig:sigLim}. Such cross-sections depend upon the cuts and details of
 the analysis in question, and we have plotted the central value of the
cross-section within bins of 20 GeV width obtained from the  fit. The CMS
analysis~\cite{Khachatryan:2016yec} 
displayed uncertainties, which are nonetheless small (even to the right-hand
side of the curve they are small). 
Fig.~\ref{fig:sigLim} also shows the 95$\%$ confidence level upper limits on the
production cross-section of a narrow resonance (we call this resonance $X$)
that decays into a two-photon 
state from ATLAS and CMS. 
The resonant di-photon channel is then assumed to be  
\begin{equation}
pp\rightarrow X+x\rightarrow \gamma\gamma+x,
\label{standard}
\end{equation} 
where $X$ is electrically neutral and can either be a spin $0$ or spin $2$
resonance, whereas $x$ is the remnant of the proton (for example, formed by
spectator quarks), which tends to remain close to the beam-line and hence
undetected. Below, we shall ignore $x$, since it is not relevant to the
phenomenology that we discuss.
There are 
quantitative differences if one takes the 
assumption of a broad resonance, but the picture is still roughly the same:
for resonances of a mass larger than 1 TeV, the cross section times branching
ratio upper limit from current experimental searches
lies somewhere between 0.1 fb and 1 fb. It is clear from the
figure that other assumptions about the resonance $X$, such as its spin, also
affect the numerical value of the bound (this is because the acceptance of the
signal changes). Assumptions about its production process: in particular,
whether it is produced by quarks or gluons\footnote{For example, the spin~2
 Randall-Sundrum graviton 
\cite{Randall:1999ee} has a well 
defined ratio of production cross-sections between $gg$ and $q \bar q$,
depending upon its mass~\cite{Allanach:2000nr}.}, also affect the signal
acceptance 
and hence the bound.

Heavy scalars are can result from models which contain two higgs
 doublets~\cite{Djouadi:2005gj}, supersymmetric extensions of little Higgs
 models~\cite{Roy:2005hg, Csaki:2008se} or extra-dimensional frameworks with
 bulk scalars~\cite{Gherghetta:2000qt}. Heavy gravitons can be
 attributed to the Kaluza Klein excitations of higher-dimensional gravity
 arising in either warped~\cite{Randall:1999ee} or
 flat~\cite{Appelquist:2000nn} geometries. The possibility of a spin 1
 particle directly decaying to di-photons is forbidden by the  Landau-Yang
 theorem~\cite{Landau:1948kw,Yang:1950rg} 
 
In some models, the heavy resonance $X$ may decay into $nn$ or $n \gamma$, where
$n$ is an additional light
particle,
may further decay into photons leading to a multi-photon\footnote{In the
  present paper, whenever we refer to multi-photon final states, we refer to
  three or more photons.} final state. 
Examples of such models include hidden
valley models~\cite{Strassler:2006im,Strassler:2006ri}, the next-to-minimal
supersymmetric standard model 
(NMSSM)~\cite{Ellwanger:2009dp} or Higgs portal
scenarios~\cite{Schabinger:2005ei}. 
There was an 8 TeV ATLAS search for a heavy  resonance decaying into
three and four photon states in Ref.~\cite{Aad:2015bua}. For
a mass of $n$ greater than 10 GeV, and a scalar $X$ of mass 600 GeV, the upper
bound on cross section times branching ratios was 1 fb. For a 
$Z'$ particle of mass 100-1000 GeV, the bound on cross-section times branching
ratio into a three-photon final state (and $n$ mass in the range 40-100 GeV)
was found to be between 35-320 fb. 
However, in the limit where $m_n\ll m_X$,  photons from $n$ will be
highly collimated, thereby creating the illusion of a di-photon final state
from the detector point of view. 
Describing angles in terms of the pseudorapidity $\eta$ and the asimuthal
angle around the beam $\phi$, the angular separation between two photons may be 
quantified by $\Delta R=\sqrt{(\Delta \eta)^2+(\Delta \phi)^2}$. 
Neglecting its mass, 
the opening angle between the two photons coming from a highly boosted
on-shell $n$ is 
\begin{equation}\Delta  R=\frac{m_n}{\sqrt{z(1-z)}p_T(n)},
\end{equation} purely from kinematics (this was
  calculated already in the context of boosted Higgs to $b \bar b$
  decays~\cite{Butterworth:2008iy}),
  where $z$ and 
 $(1-z)$ are  the momentum fractions of the photons\footnote{The decay is
   strongly peaked towards the minimum opening angle $\Delta R=2
   m_n/p_T$~\protect\cite{Chala:2015cev}.}. 
 Thus,
\begin{equation}
\Delta R = \frac{m_n}{M_X}\frac{2 \cosh \eta(n)}{\sqrt{z(1-z)}}. \label{dRest}
\end{equation}
In the limit $m_n/M_X\rightarrow 0$, $\Delta
R\rightarrow 0$ and the two photons from $n$ are 
collinear, appearing as one photon; thus several possible interpretations
can be ascribed to an apparent di-photon signal.

Below, we shall examine
the phenomenology of apparent $\gamma 
\gamma$ resonances, ignoring backgrounds. For this to be a good approximation,
we require that the background is small compared to the signal cross-section. 
Fig.~\ref{fig:sigLim} shows that for $m_X\simgt 1200$ GeV, 
there is parameter space where this is the case, i.e.\ where $\sigma(pp
\rightarrow X)\ BR(X \rightarrow \gamma \gamma)$ is well above the background
but below the current experimental limits.
 The scenarios corresponding to different spins of $X$ and $n$  may be
 characterised by distributions of $\Delta \eta$ between 
 the apparent di-photon states.
 Differences in the predicted $\Delta \eta$ distributions allows us to
 estimate the minimum number of events 
 needed to discriminate between the different cases. In the event that the
 mass of 
 the intermediate state $n$ is not {\em too}\/ small, such that the photons from
 it 
 can often be resolved, the multi-photon topology can be distinguished from the
 di-photon topology using the substructure of photon
 jets~\cite{Ellis:2012zp,Ellis:2012sd}. However, in the limit
 $m_n/m_X\rightarrow 
 0$, it is hard to resolve the photons from $n$. 

There has been earlier work on heavy $X$ spin discrimination in a truly
di-photon final state:
telling spin~0 from spin~2~\cite{Kumar:2011yta,Alves:2012fb,Pankov:2012oxa}.
However, our paper goes beyond these: we consider multi-photon cases which
only appear to be di-photon cases at the first glance. 

It will be useful for us to categorise models' signatures into 2 classes:
the first is multi-photon signals, where $m_n$ is large enough for the
photons (from $n$) to be detected by different cells of the electromagnetic
calorimeter, but small enough so that they produce the illusion of a single
photon.
The other category includes both the standard di-photon topology and the
multi-photon topology in the limit $m_n/M_X\rightarrow 0$. Each apparent photon
lies within a single cell of the electromagnetic calorimeter.
These cases might be discriminated by photon jet
substructure properties. We shall use substructure variables to identify the
fundamental nature of the 
topology  and conventional kinematic
variables to distinguish the different spin possibilities in each case. 

The paper is organised as follows: in section~\ref{modeldescription} we set up
extensions to the SM Lagrangian which can predict heavy di-photon
or multi-photon resonances. The finite photon resolution of the detector
is discussed in section~\ref{photonsize}.
In Section~\ref{photonjets},
isolation criteria are removed and photon-jets are
adopted. Substructure and kinematic observables are then used 
to distinguish the different scenarios. In section~\ref{kl} we
introduce the statistics which tell us how many measured signal events will be
required to discriminate one set of spins from another, whereas we cover how
one can constrain the mass of the intermediate particle $n$ in
section~\ref{mas}. 
We conclude 
in section~\ref{conclusion}. Appendix~\ref{sigmod} contains some details about model
parameters.

\section{Model description}
\label{modeldescription}
In this section we describe the minimal addition to the SM Lagrangian which
can give  rise to heavy resonant final states made of photons. We make no
claims of generality: various couplings not relevant for our final state 
or production will be neglected. However, we shall insist on SM gauge
invariance. 
Beginning with the di-photon final state, a minimal extension 
involves the introduction of a SM singlet heavy resonance $X$.  
We assume that any couplings of new particles such as the $X$ (and the $n$, to
be introduced later) to Higgs
fields or $W^\pm, Z^0$ bosons are negligible. 
Eq.~\ref{Lagrangian1} gives an effective field theoretic interaction
Lagrangian for the 
coupling of $X$ to a pair of photons, when $X$ is a scalar (first line) or  a
graviton (second line). 
 \begin{eqnarray}
	\mathcal{L}_{X={\rm spin~0}}^{int} &=&    - \eta_{GX}\frac{1}{4} G_{\mu \nu}^a
                                   G^{\mu \nu a} X  - \eta_{\gamma X} \frac{1}{4}
                                   F_{\mu \nu} F^{\mu \nu} X, \nonumber\\
	\mathcal{L}_{X={\rm spin~2}}^{int} &=&    -\eta_{T\psi X}T_{fermion}^{\alpha\beta}
                                   X_{\alpha\beta}-\eta_{TGX} T_{gluon}^{\alpha\beta}
                                   X_{\alpha\beta}  - \eta_{T\gamma X} 
T_{photon}^{\alpha\beta} X_{\alpha\beta}.
	\label{Lagrangian1}
\end{eqnarray}
where  $T^{\alpha\beta}_i$ is the stress-energy tensor for the  field $i$
and the $\eta_j$ are effective couplings of mass dimension -1.
$F_{\mu \nu}$ is the field strength tensor of the photon (this may be obtained
in a SM invariant way from a coupling involving the field strength tensor of
the hypercharge gauge boson),
whereas $G_{\mu 
  \nu}^a$ is the field strength tensor of a gluon of adjoint colour index $a
\in \{ 1,\ \ldots , 8 \}$.
As noted earlier, the direct decay of a vector boson into two photons is
forbidden by the 
Landau-Yang theorem~\cite{Landau:1948kw,Yang:1950rg}. Since $X$ is assumed to
be a SM singlet, there are no couplings to SM fermions, which are in
non-trivial chiral representations when it is a scalar. 

The presence of an additional light scalar SM singlet in the theory ($n$),
with masses such 
that $m_n<m_X$, opens up another decay mode: $X \rightarrow nn$. 
Lagrangian terms for these interactions are 
\begin{equation}
	\mathcal{L}^{int}_{X={\rm spin~0},n} =    
                                   - \frac{1}{2}A_{Xnn} X n n , \qquad
\mathcal{L}^{int}_{X={\rm spin~2},n} = 
-\eta_{TnX} X_{\alpha \beta} T^{\alpha \beta}_n,
\end{equation}
where $A_{Xnn}$ has mass dimension 1.
$n$ may
further decay into a pair of  
photons leading 
to a multi-photon final state through
a Lagrangian term
\begin{equation}
\mathcal{L}^{int}_{n\gamma \gamma}=-\frac{1}{4}\eta_{n\gamma\gamma} F_{\mu \nu} F^{\mu
  \nu} n. 
\end{equation}
Although we assume that $n$ is electrically neutral, it may decay to two
photons through a loop-level process (as is the case for the Standard Model
Higgs boson, for instance). 
Alternatively, if $X$ is a spin 1 particle, it could be produced by
quarks in the proton and then decay into $n \gamma$. 
The Lagrangian terms would be
\begin{equation}
\mathcal{L}^{int}_{X\,={\rm spin~1},n}= -
                                     (\lambda_{\bar q X q} \bar q_R \gamma_\mu
                                     X^\mu q_R
 + \lambda_{\bar Q X Q} \bar Q_L \gamma_\mu X^\mu Q_L + H.c.)
-\frac{1}{4}\eta_{nX\gamma}
                                     n{\tilde X}_{\mu\nu}F^{\mu\nu},
\end{equation}
where  $\lambda_i$ are  dimensionless
couplings, $q_R$ is a right-handed quark, $Q_L$ is a left-handed quark doublet
and ${\tilde X}_{\mu \nu}=\partial _\mu
X_\nu - \partial_\mu X_\nu$. 
The decay $X_{spin=1}\rightarrow n \gamma$ would
have to be a loop-level process, as explicitly exemplified in
Ref.~\cite{Chala:2015cev}, since electromagnetic gauge invariance 
forbids it at tree level. 
A spin 1 particle may not decay into two identical spin
0 bosons due to Bose symmetry: the daughters must be symmetric under
interchange, meaning they must have even orbital angular momentum $L$. 
Then it is impossible to conserve total angular momentum $J$ since the initial
state has $J=1$ and the final state has $J$ even.

For scalar $n$ then, we have a potential
four photon final state if $X$ is spin~0 or spin~2 and a potential three photon
final state if $X$ is spin~1 as shown in Eq.~\ref{collimated}:
\begin{eqnarray}
p~p\rightarrow X_{spin=0,2}&\rightarrow& n~n\rightarrow \gamma\gamma +\gamma\gamma \nonumber\\
	p~p\rightarrow X_{spin=1}&\rightarrow& n~\gamma\rightarrow \gamma\gamma+\gamma
\label{collimated}
\end{eqnarray}
If the mass of the intermediate scalar $n$ is such that $m_n \ll m_X$, its
decay products are highly collimated because the $n$ is highly boosted. 
It thereby results in a photon pair resembling a single photon final state.
This opens up a range of possibilities with regards to the interpretation of the
apparent di-photon channel. 
Above, we have assumed the intermediate particle $n$ to be a
scalar while considering different possibilities for the spin of $X$. 
Table~\ref{possibilities} gives possible spin combinations for the heavy
resonance $X$
and the intermediate particle $n$ leading to a final state made of photons. 
The third column gives the number of photons for each topology,
grouped in terms of collimated photons that may experimentally resemble a single
photon in the $m_n / m_X \rightarrow 0$ limit. 
The 
spin~1 $X$ example was already proposed as a possible
explanation~\cite{Chala:2015cev} for a putative 
750 GeV apparent di-photon excess measured by the LHC experiments (this
subsequently turned out to be a statistical fluctuation).
\begin{table}
	\begin{center}
		\begin{tabular}{|c|c|c|} 
			\hline
			Spin of $X$&Spin of $n$ &Number of
			photons\\ 
			\hline
			\multirow{2}{*}{0}& 0&$\gamma\gamma$+$\gamma\gamma$\\
                                   & 2  &$\gamma\gamma$+$\gamma\gamma$\\ 
			\hline
			\multirow{2}{*}{1}& 0 &$\gamma$+$\gamma\gamma$\\
& 2 &$\gamma$+$\gamma\gamma$\\
			\hline
			\multirow{2}{*}{2}&
                                                 0&$\gamma\gamma$+$\gamma\gamma$\\&
                                            2&$\gamma\gamma$+$\gamma\gamma$\\
			\hline
		\end{tabular}
	\end{center}
	\caption{Different possibilities for spin assignments leading to 
          an apparent di-photon state from other multi-photon
		final states. The one or two photon
		states   have been grouped into terms which may only be
                resolved as one
		photon when $m_n/m_X$ is small.}
	\label{possibilities}
\end{table}

In this work, we shall focus  on the case where $n$ is a scalar. 
However, the techniques developed in this paper 
can be extended to cases where $n$ is spin 2 as well (but not spin 1, since
$n \rightarrow \gamma \gamma$ would then be forbidden by the Landau-Yang
theorem).  
In the next section
we 
will describe the scenario under which the process in Eq.~\ref{collimated} can
mimic a truly di-photon signal.

\section{The size of a photon}
\label{photonsize}
In a collider environment, any given process can be characterised by a given
 combination of final states. These final states correspond to different
 combinations of  
photons, leptons (electrons and muons), jets and missing energy.
They can be distinguished by the energy deposited by them in different
sections of the detector.  
In a typical high energy QCD jet, most of the final state particles (roughly
2/3) are  charged pions whereas 
neutral pions make up much of the remaining 1/3~\cite{Ellis:2012zp}. The
constituents of a jet 
primarily deposit their energy in the 
hadronic calorimeter
(HCAL) while the $\pi^0\rightarrow 2\gamma$ decay of a neutral pion ensures
that it shows up in the electromagnetic calorimeter (ECAL). Thus most of the constituents of the jet
pass through the ECAL and deposit their energy in the HCAL\@. Photons and
electrons deposit their energy in the ECAL, on the other hand. They can be
distinguished by  mapping the energy deposition to the tracker (which precedes
the calorimeters). Apart from the tracker, electrons and photons are
similar in appearance, from a detector point of view. Muons are detected by
the muon spectrometer on the outside of the experiment.

We shall now go on to discuss the relevant parts of
the detectors and experimental analyses. 
The actual construction and workings
of the detector are of 
course much more detailed than we, outside of the experimental collaborations,
have tools for dealing with. We therefore characterise the cuts and detector
response 
in {\em in broad brush strokes}. With this in mind,
the experimental sensitivity to detect a single photon is subject to
the following two criteria:\\  
(a) \textbf{Dimensions of the ECAL cells}:  The ATLAS and CMS
detectors have slightly different dimensions for the ECAL cells. ATLAS has
a slightly coarser granularity with a 
crystal size of $(0.0256,\ 0.0254)$ in $(\eta,\phi)$. 
In
comparison, CMS has a
granularity of $(0.0174,\ 0.0174)$ in $(\eta,\ \phi)$. 
CMS and ATLAS have
a layer in their electromagnetic 
calorimeters with finer $\eta$ segmentation (in ATLAS, this is called `layer
1') but worse $\phi$ segmentation, which could also be employed in 
analyses looking for resonances into multi-photon final states. The level
of ECAL modelling including 
this layer is beyond the scope of this paper, and so we do not
       discuss it further. However, we bear in mind that information from the
       layer 1 may be used {\em in addition}\/ to the techniques
       developed in this paper. Any estimates of sensitivity (which come
       later) are therefore conservative in the sense that additional
       information from layer 1 could improve the sensitivity.
High energy photons will tend to shower in the ECAL: this is
taken into account by clustering the
cells into cones of size $R_{cone}=\Delta R=0.1$.  
Thus if two high energy signal photons are separated a distance $\Delta
R<R_{cone}$, they are  
typically {\em not}\/ considered to be resolved by the ECAL since it could be
a single photon that is simply showering.\\  
(b) \textbf{Photon isolation}: In ATLAS and in CMS, a photon is considered to
 be isolated 
if  
the magnitude of the vector sum of the transverse momenta ($p_T$) of all
objects with $\Delta R \in [R_{cone},\ 0.4]$ is less than 10$\%$ of its $p_T$.
Qualitatively, this corresponds to the requirement that most of the energy is carried by the photon around which the cone is constructed.
This criterion is required in order to distinguish a hard photon from a photon
from a $\pi^0$ decay. 

However, it is possible that certain {\em signal}\/ topologies may give rise
final state 
photons that are separated by a distance $\Delta R \in [R_{cone},\ 0.4]$. 
For instance, consider the process given in Eq.~\ref{collimated}. The particle
$X$ can either be a scalar or a graviton. For concreteness, let us assume that
$n$ is a scalar. 
In this case, a four photon final state
resulting from $X \rightarrow nn \rightarrow \gamma \gamma + \gamma \gamma$
would appear to be a di-photon final state.   
However, as $m_n$ increases, eventually $\Delta R>0.4$ and the number of
resolved final state photons will increase. 
Similar arguments hold for the case  where particle $X$ is a spin~1 state.
For a given mass of $n$, the eventual number of detected, isolated and
resolved photons depends on the granularity of the detector and is expected to
be slightly different for both the CMS and ATLAS. 

To approximate the
acceptance and efficiency of the detectors for our signal process, we perform
a Monte-Carlo simulation using the following steps: 
\begin{itemize}
	\item The matrix element for our signal process is generated in
          {\tt{MadGraph5 aMC@NLO}}~\cite{Alwall:2014hca} by generating the
          Feynman rules for the process with
          {\tt{FEYNRULES}}~\cite{Christensen:2008py}. We set $\eta_i
          =\mathcal{O} 
          (\textrm{20~TeV})^{-1}$ as specified in Appendix~\ref{sigmod}, 
          $A_{Xnn}=M_X/100$
          and $\lambda_i=0.5$ in the model file. {\tt
            MadGraph5} 
          then calculates the width of the $X$: $\Gamma_X\sim 1-2$ GeV
          depending on the model, so the
          heavy resonance is narrow\footnote{The light resonance is also
            narrow, since 
          $\Gamma_n = m_n^3 |\eta_{n \gamma \gamma}|^2 / (64 \pi)$.}.
          Events are generated at 13 TeV centre of
          mass energy using the 
          {\tt{NNLO1}}~\cite{Ball:2012cx} parton distribution functions.
        \item For showering and hadronisation, we use {\tt{PYTHIA
              8.2.1}}~\cite{Sjostrand:2007gs}. The set of final state
          particles is then passed through the {\tt{DELPHES 
              3.3.2}} detector simulator~\cite{deFavereau:2013fsa}.
\end{itemize}
We use the {\tt{DELPHES 3.3.2}} isolation module for photons and we impose a
minimum $p_T$ requirement of $100~$GeV on each isolated photon. 
\begin{figure}[h]
\begin{center}
	\begin{tabular}{cc}
\includegraphics[width=0.49\textwidth]{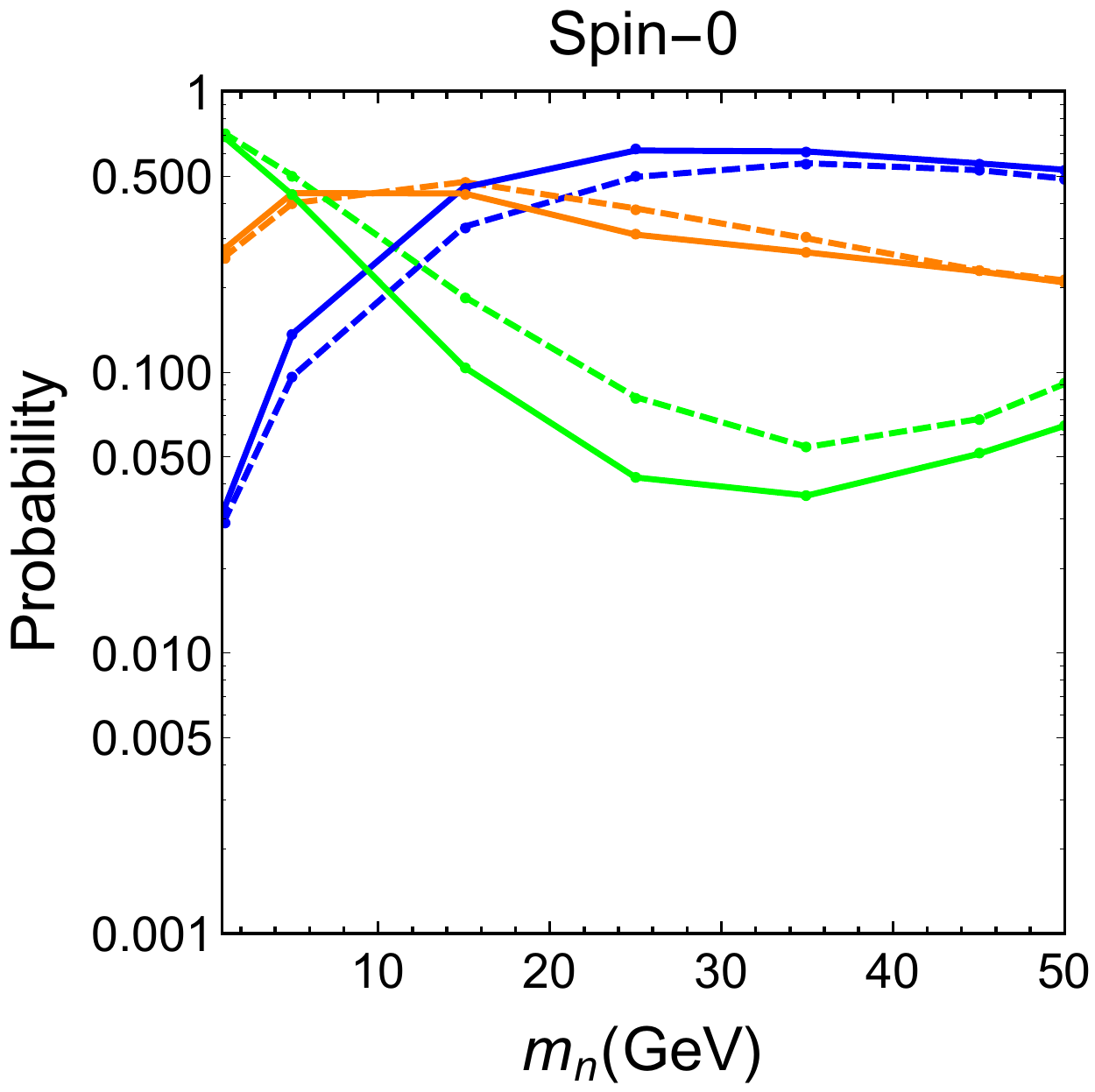}&\includegraphics[width=0.49\textwidth]{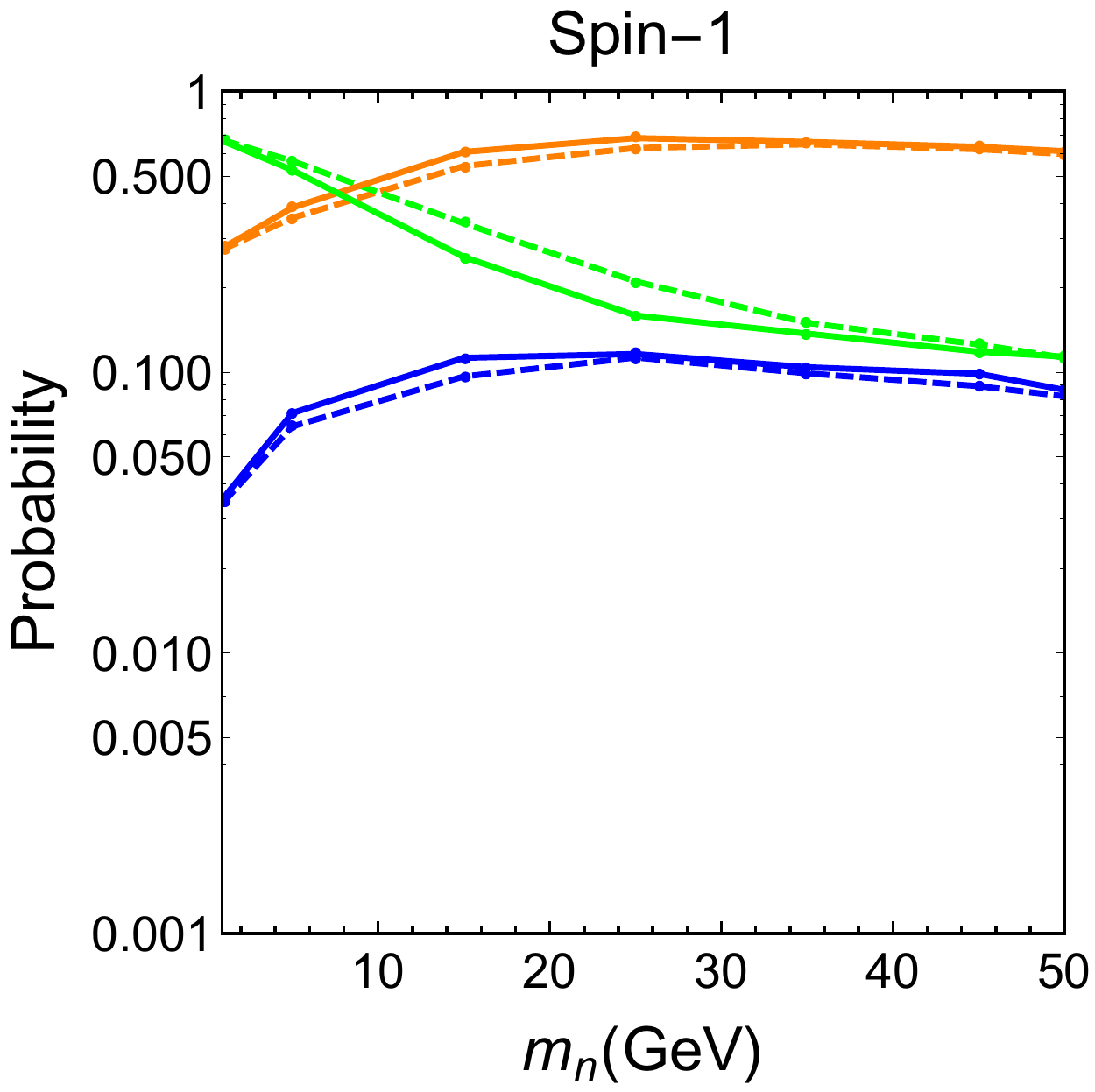}\\
\includegraphics[width=0.49\textwidth]{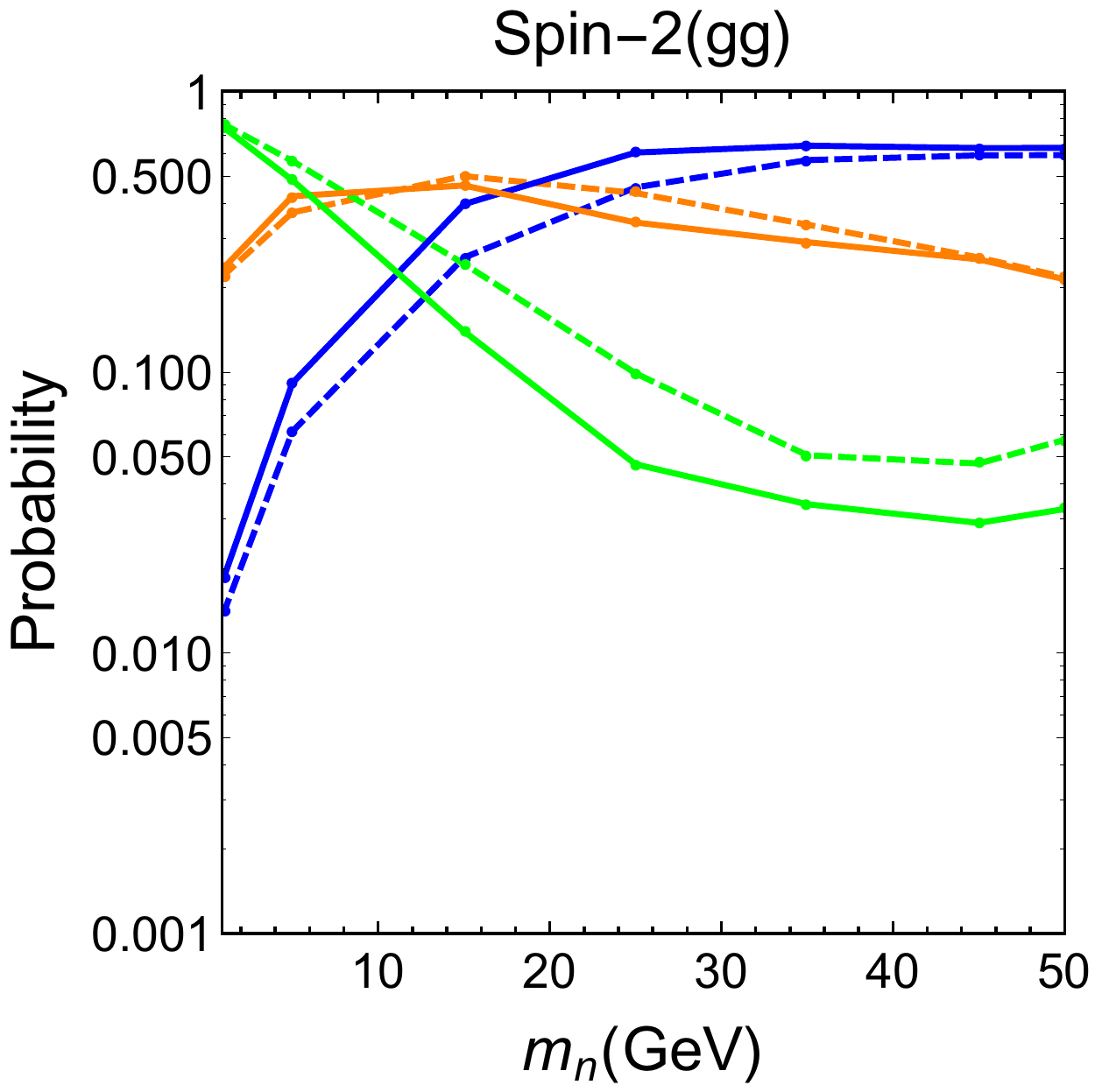}&\includegraphics[width=0.49\textwidth]{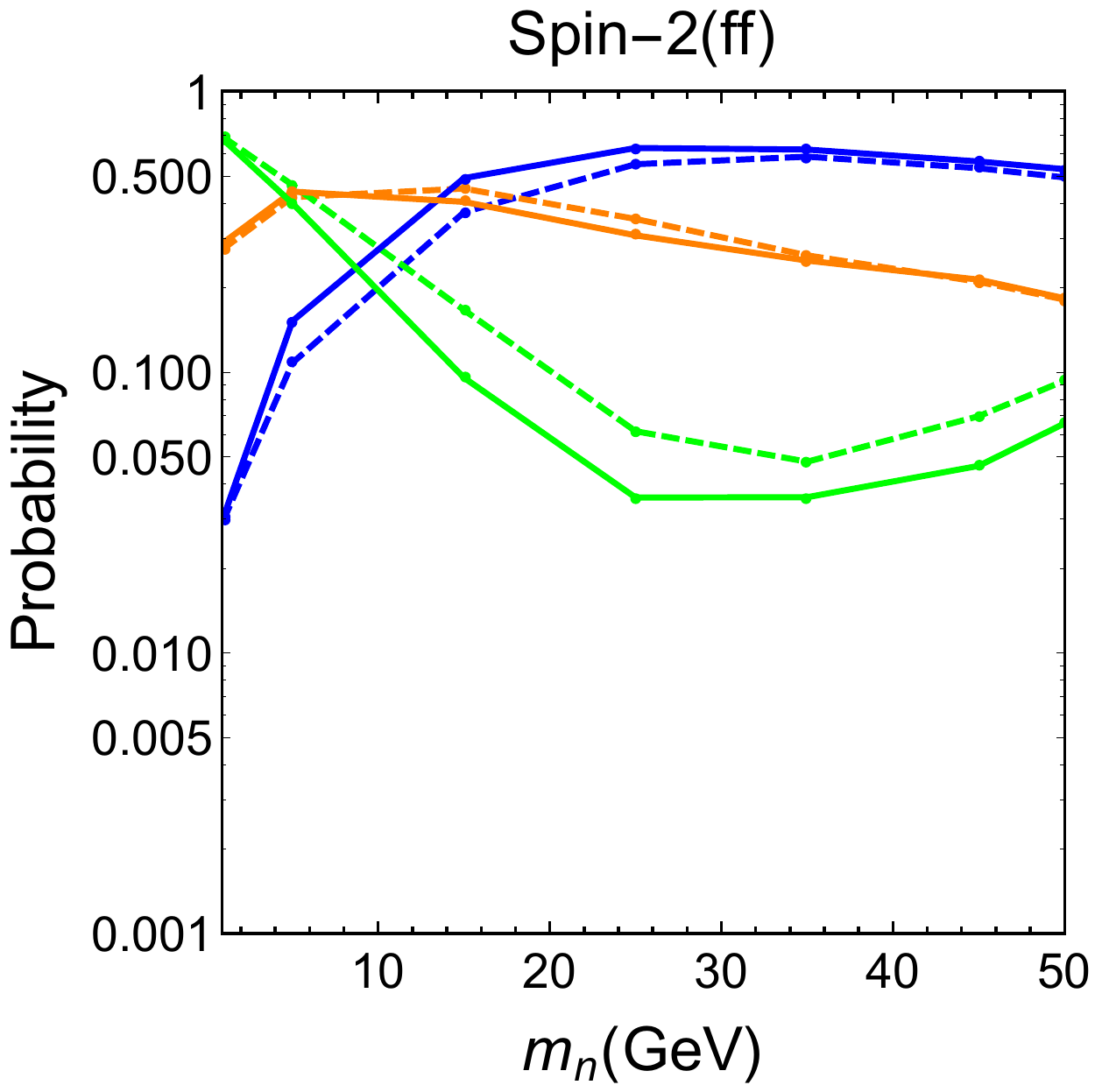}\\
\end{tabular}
	\caption{Probabilities of detecting different numbers of isolated,
          resolved photons for
          a 1200 GeV
          $X \rightarrow$ multi-photon decay
          as a function of $m_n$, the mass of the intermediate particle. We
          show the probabilities for 
0 (blue), 1 (orange) or 2 (green) photons for each $X$ produced. The
probabilities for detecting 3 or 4 isolated, resolved photons for the signal
are very small for this range of $m_n$ and are not shown.
Solid lines correspond to CMS, and dashed lines to ATLAS. }
	\label{comparison}
	\end{center}
\end{figure}

Figure~\ref{comparison} shows the probabilities of detecting the different
number of detected, resolved, isolated photons in the 
final state  for a produced $X$  for ATLAS (dashed) and CMS
(solid).  If $p_T(\gamma)<10$ GeV or $|\eta(\gamma)|>2.5$, {\tt DELPHES}\/
records a 
zero efficiency for the photon, and it is added to the `0 photon' line. In the
rest of the detector, {\tt DELPHES}\/ assigns between a 85$\%$ and a 95$\%$
weight for the photon (the difference from 100$\%$ is also added to the `0
photon' line in the figure). A few of the simulated photons from the $X$
additionally fail the 
$p_T>100$ GeV cut: these are not counted in the figure, and so the curves do
not add exactly to 1. 

The probabilities are
shown for different possibilities of the spin of $X$, as shown by the header
in each case. 
The bottom row corresponds to spin 2 when it is produced by $gg$ fusion
(left) and $\bar qq$ annihilation (right). 
Spin 1 corresponds to $X\rightarrow n \gamma \rightarrow \gamma \gamma +
\gamma$, whereas the other cases all correspond to a $X \rightarrow nn
\rightarrow \gamma \gamma + \gamma \gamma$ decay chain.
The effective number of 
detected photons can be reduced by them not appearing in the fiducial volume
of the detector (i.e.\ $|\eta(\gamma)|<2.5$), or by them not being isolated
(in which case both photons are rejected) or
resolved (in which they count as one photon). We note that for each spin case,
in the low $m_n$ limit, the $X$ is most likely to be seen as two resolved,
isolated photons because each photon pair is highly collimated.
 
\begin{figure}[h]
	\begin{center}
			\includegraphics[width=0.6\textwidth]{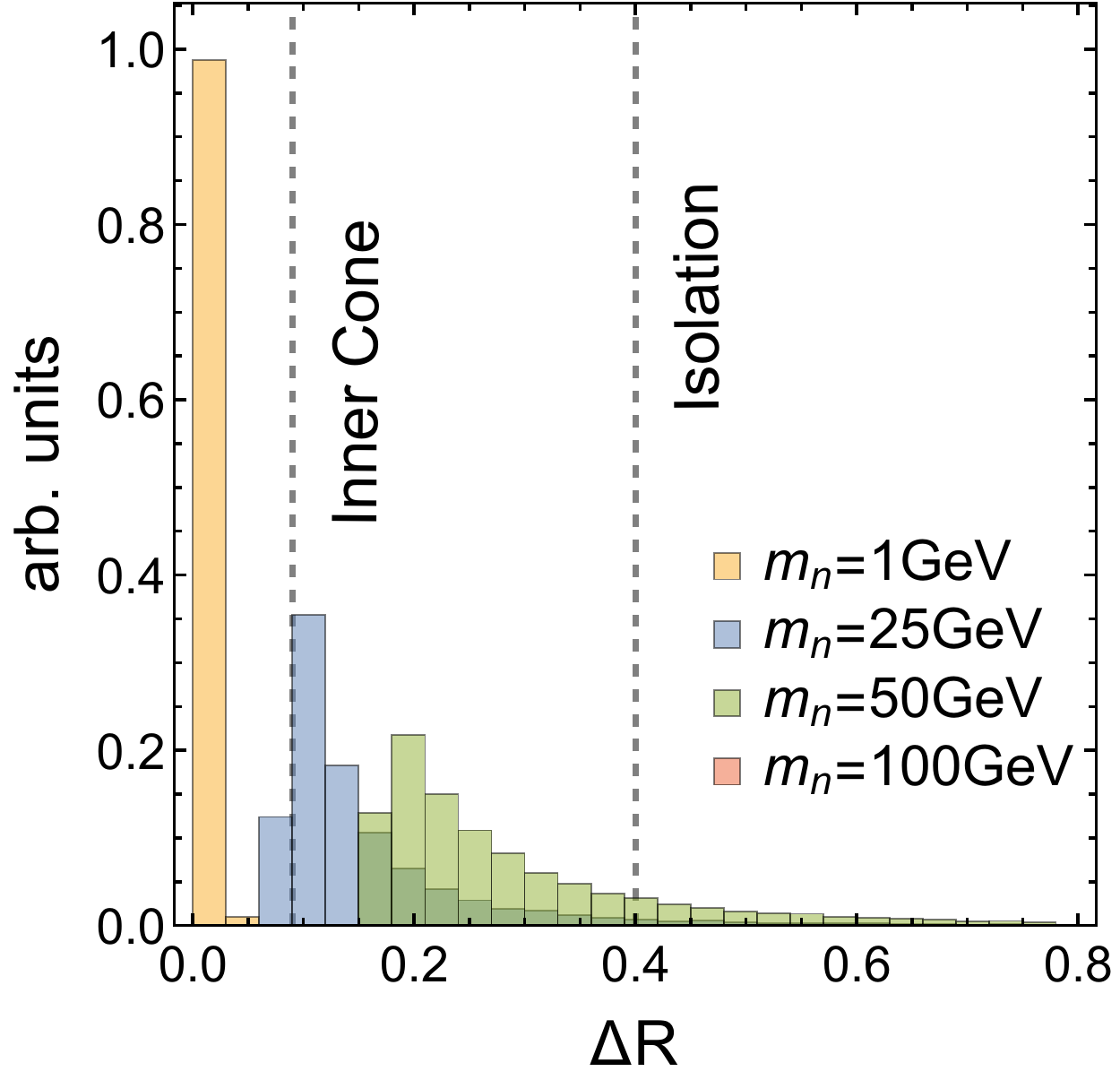}\\
		\caption{$\Delta R$ distribution for photon pairs originating
                  from $n\rightarrow  \gamma \gamma$  for
                  different values of $m_n$. Photon pairs to the left hand
                  side of the `ECAL Prescription' line are considered to be
                  one photon, whereas those between the ECAL prescription and
                  the `Isolation' line are rejected because of the photon
                  isolation criteria.} 
		\label{isolation}
	\end{center}
\end{figure}

We note first that the probability for detecting 0, 1 or 2 resolved, isolated
photons for the spin 2 case does not depend much on whether it is produced by a
hard $gg$ collision or a hard $\bar q q$ collision. 
An interesting trend is observed for the spin 0 and spin 2 cases, where the
two photon probability has a minimum at $m_n\approx 40$ GeV. 
At $m_n=40$ GeV, the photon pair from an $n$ are often separated by 
$\Delta R \in [R_{cone},\ 0.4]$ and fail the isolation criterion
because the two
photons have similar $p_T$.  Fig.~\ref{isolation} gives the distribution of
$\Delta R$ between the photon pair coming 
from $n$ as a function of its mass, and illustrates the preceding point. For
light masses ($m_n=1 $ GeV) it is 
clear that both signal photons are 
within 
$\Delta R<R_{cone}$. For intermediate masses $m_n \in \{25,\ 50\}$ GeV, most
photons are within $\Delta R \in [R_{cone},\ 0.4]$, whereas for $m_n=100$ GeV,
a good fraction are already isolated photons, having $\Delta R > 0.4$.   
Using an estimate $m_n \sim M_X \Delta R / 4$ from Eq.~\ref{dRest}, 
we deduce that events with four 
isolated signal photons are expected to be evident only in the $m_n
 \gsim 120$ GeV region for $M_X=1200$ GeV.

The spin 1 case in comparison, has a significantly lower zero photon rate for 
$m_n < 50$ GeV,  as the process is characterised by a single photon and two
collimated photons. Thus, unless the single photon is lost in the barrel or lost because of tagging efficiency, it
will be recorded even if the collimated photons fail the isolation
criterion.

\section{Photon Jets} 
\label{photonjets}
Since we wish to describe collimated and non-isolated photons in more detail
(since, as the previous section shows, these are the main mechanisms by which
signal photons are lost), we follow refs.~\cite{Ellis:2012sd,Ellis:2012zp} and
define photon-jets. 
For this, we relax the
isolation 
criteria and work with the detector objects, i.e.\ the
calorimetric and track four vectors. 
The calorimetric four vectors for each event are required to satisfy the
following acceptance criteria: 
\begin{equation}
	E_{ECAL}>0.1 \text{~GeV}\quad\;, \quad E_{HCAL}>0.5 \text{~GeV},
\end{equation}
while only tracks with $p_T>2$ GeV are accepted.
These calorimetric and track four vectors are clustered using {\tt{FASTJET
    3.1.3}}~\cite{Cacciari:2011ma} using the anti-$k_T$~\cite{Cacciari:2008gp}
clustering algorithm with $R=0.4$. The 
tracks' four vectors are scaled by a small number and are called `ghost
tracks': their directions are well defined, but this effectively scales down
their energies 
to negligible levels to avoid
over counting them (the energies are then defined from the calorimetric
deposits). 
The photon jet size $R=0.4$ is
chosen to coincide with the isolation separation of the  photon described in
Section~\ref{photonsize}. 
The anti-$k_T$ clustering algorithm ensures that the jets are
well defined cones (similar to the isolation cone) and clustered around a
hard momentum four vector, which lies at the centre of the cone. Thus for our
signal 
events, the jets are constructed around the photon(s). These typically have a
large $p_T$, since they are produced from a massive resonance. 

Since these jets are constructed out of the calorimetric (and ghost track)
four vectors, they constitute a starting point for our analysis.  
At this stage, while a QCD-jet (typically initiated by a quark or gluon) is on
the same footing as a photon jet, they can be discriminated from each
other\footnote{Here we have not implemented such cuts, since we only simulated
signal.} by
analysing different observables:
\begin{itemize}
	\item \textbf{Invariant mass cut:} We would demand the invariant mass
          of the 
          two leading photon jets to be close  to the mass of the observed
          resonance, reducing continuum backgrounds.
	\item \textbf{Tracks:} QCD jets are composed of a large number of
          charged mesons which display tracks in the tracker before their
          energy is deposited in the 
          calorimeter\footnote{A gluon initiated jet typically has a larger
 track multiplicity than a quark initiated
            jet.}~\cite{Bhattacherjee:2015psa}. The track 
          distribution for a QCD jet typically peaks at higher values  of
          the number of tracks compared 
          to a photon jet which peaks at zero tracks. 
	\item \textbf{Logarithmic hadronic energy fraction ({$\log\theta_J $}):}
          This variable 
          is a measure of the hadronic energy fraction of the jet. 
For a photon jet most of the energy is carried by the hard photon(s). As a
result, this jet will deposit almost all of its energy into the ECAL, which is
in 
stark contrast with a QCD jet. This can be quantified by constructing the
following substructure observable~\cite{Ellis:2012sd,Ellis:2012zp}:
	\begin{equation}
	\theta_J=\frac{1}{E^{total}}\sum_iE_i^{HCAL},
	\end{equation}
where $E^{total}$ is the total energy in the jet deposited in the HCAL plus
that deposited in the ECAL, whereas $E_i^{HCAL}$ is the energy of each jet
sub-object $i$ that is deposited in the HCAL.
$\log(\theta_J)$ is large and negative for a photon jet, while it peaks
        close to $\log[2/3]=-0.2$ for a QCD jet, since charged pions constitute
        around $(2/3)$ of the jet constituents.  We would require the leading
        jet to have $\log(\theta_J)<-0.5$, corresponding to very low hadronic
        activity. 
\end{itemize}
 Under these cuts, the $QCD$ fake rate should reduce to less than
 $10^{-5}$~\cite{Ellis:2012sd,Ellis:2012zp}.
Removing photon isolation and instead describing the event in terms of photon
jets is advantageous because it  
helps discriminate the standard di-photon decay in Eq.~\ref{standard} from the
decay to more than two photons in Eq.~\ref{collimated}. 
However, it still fails in the limit $m_n/M_X \rightarrow 0$, as we shall see
later. 
Taking photon jets as a starting point,
we shall devise strategies where we may
discern the nature of the topology
and glean information about the spins of the particles involved.  

\subsection{Nature of the topology}
\label{variables}

\begin{table}\begin{center}
\begin{tabular}{|c|c|}\hline
Model & Process \\ \hline
$S2$ & $pp \rightarrow S \rightarrow \gamma \gamma$ \\
$S4$ & $pp \rightarrow S \rightarrow nn \rightarrow \gamma
       \gamma+\gamma\gamma$ \\ 
$V3$ & $pp \rightarrow Z' \rightarrow n \gamma \rightarrow \gamma + \gamma
       \gamma$ \\
$G2_{ff}$ & $q \bar q \rightarrow G \rightarrow \gamma \gamma$ \\ 
$G4_{gg}$ & $gg \rightarrow G \rightarrow nn \rightarrow \gamma \gamma+\gamma
            \gamma$ \\
$G4_{ff}$ & $\bar q q \rightarrow G \rightarrow nn \rightarrow \gamma \gamma+\gamma
            \gamma$ \\
\hline
\end{tabular}
\end{center}
\caption{\label{tab:cases} Cases to discriminate with a scalar $n$ and a heavy
resonance which is: scalar ($S$), spin 1 ($Z'$) or spin 2 ($G$). We have
listed the main signal processes to discriminate between in the second column,
ignoring any proton remnants. The notation used for a given model is $Xk$:
$X=S,V,G$ labels the spin of the resonance and $k$ denotes the number of
signal photons at the parton level in the final state.} 
\end{table}
In this section we identify variables that aid  in identifying the
topology of the signal process and the spin of $X$.  We begin by listing
different cases we would 
like to discriminate between in Table~\ref{tab:cases}. 
In the event of an observed excess in an
apparent 
di-photon final state, we would relax the isolation criteria and  
define photon jets. 
Analysing the photon jets' substructure will help measure the number of hard
photons within each jet. 
The difference in substructure for a photon jet with a single hard photon as
opposed to several hard photons can be quantified
by~\cite{Ellis:2012sd,Ellis:2012zp}:  
\begin{equation}
\lambda_J=\log\left(1-\frac{p_{T_L}}{p_{T_J}}\right).
\label{lambdaj}
\end{equation}
This can be understood as follows:
\begin{itemize}
	\item Hard photon jets are re-clustered into sub-jets.
	\item $p_{T_L}$ denotes the $p_T$ of the leading sub-jet (i.e.\ the
          sub-jet with the largest $p_T$) within the jet in question, whilst
          $p_{T_J}$ is the $p_T$ of the parent jet. 
	\item For a `single pronged' photon jet, $p_{T_L}\sim p_{T_J} $. Thus
          $\lambda_J$ is negative, with a large magnitude.
	\item For a double-prong photon jet, $p_{T_L}<p_{T_J}$, 
          resulting in $\lambda_J$ closer to zero than the single pronged
          jets. We expect a peak where $p_T(n)$ is shared equally between the
          two photons, i.e.\ $p_{T_L}/p_{T_J}=1/2$, or {$\lambda_J=-0.3$}.
\end{itemize}
There exist other substructure variables one could use in place of
$\lambda_J$, such as 
$N-$Subjettiness~\cite{Thaler:2010tr,Thaler:2011gf} or energy
correlations~\cite{Larkoski:2013eya} which are a measure of how pronged a jet
is. Here, we prefer to use 
$\lambda_J$ because it is particularly easily implemented and understood, and is
robust in the presence of pile-up~\cite{chakraborty}.  

\begin{figure}
	\begin{tabular}{c}
		\includegraphics[width=16cm]{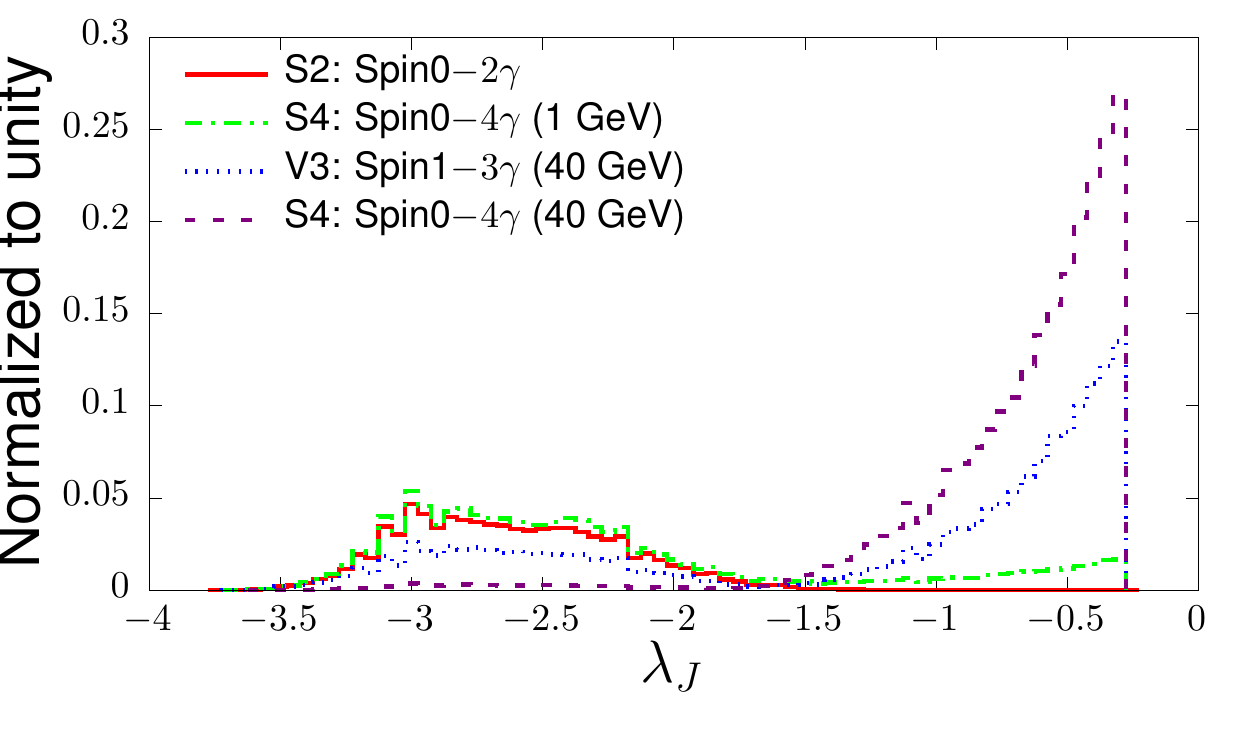}
	\end{tabular}
	\caption{Distribution of $\lambda_J$ for  $S2$ 
          and some multi-photon topologies $S4$ for $m_n=1$ GeV and $V3$ and $S4$ for $m_n=40$ GeV in the ATLAS
          detector. Double
photon jets dominantly appear at $\lambda_J \sim -0.3$.
If a single hard photon in a jet radiates, it often appears in the bump
$\lambda_J \in [-3.5,-2]$, but there is a possibility for the photon jet to
really only contain one photon: here, $\lambda_J$ is strictly minus infinity.
We do
not show such events here on the figure, but they will count toward model
discrimination.}
\label{collimatedvsstandard}
\end{figure}
Fig.~\ref{collimatedvsstandard} shows the distribution of 
$\lambda_J$ for the di-photon heavy resonance S2 (solid) and a
multi\footnote{In this 
  article, we refer to three or more hard signal photons as a multi-photon
  state.}-photon 
S4 topology 
$m_n=1$ GeV (dot-dashed). 
It is evident from the figure that
the $\lambda_J$ distribution is similar
for the two cases, since they both peak at highly negative $\lambda_J$.
This can be attributed to the fact that for such low masses of $n$ in S4, the
decay photons 
are highly collimated with $\Delta R < R_{\rm cell}$. They therefore
should resemble a single photon.
However, the appearance of a small bump like feature on the right of the plot
for $m_n=1$ GeV S4 
is interesting and unexpected {\em prima facie}\/ since the opening angle
between the photons in this case is less than the dimensions of an ECAL
cell. However, this is explained by the fact that the energy of a photon
becomes smeared around the 
cell where it deposits most of its energy. When a single (or two closely spaced
photons) hit the centre of the cell, the smearing is almost identical for both
cases. However, there exist a small fraction of cases for the collimated S4
topologies, where the two photons hit a cell near its edge such that they get
deposited in adjacent cells, leading to the small double-pronged jet peak at
{$\lambda_J=-0.3$}. 
One would require both good statistics and a very good modelling of the ECAL in
order to be able to 
claim discrimination of the two cases S2 and S4 (1 GeV), and for now we assume
that they will not be. On the other hand, by the time that $m_n$ reaches 40
GeV, the 
multi-photon topologies V3 and S4 are easily discriminated from S2, due to the
large double-photon peak at {$\lambda_J=-0.3$}. They should also be
easily discriminated from each 
other since V3 has a characteristic double peak due to its
$\gamma+\gamma\gamma$ 
topology. 

\begin{figure}
	\begin{tabular}{c}
		\includegraphics[width=0.8 \textwidth]{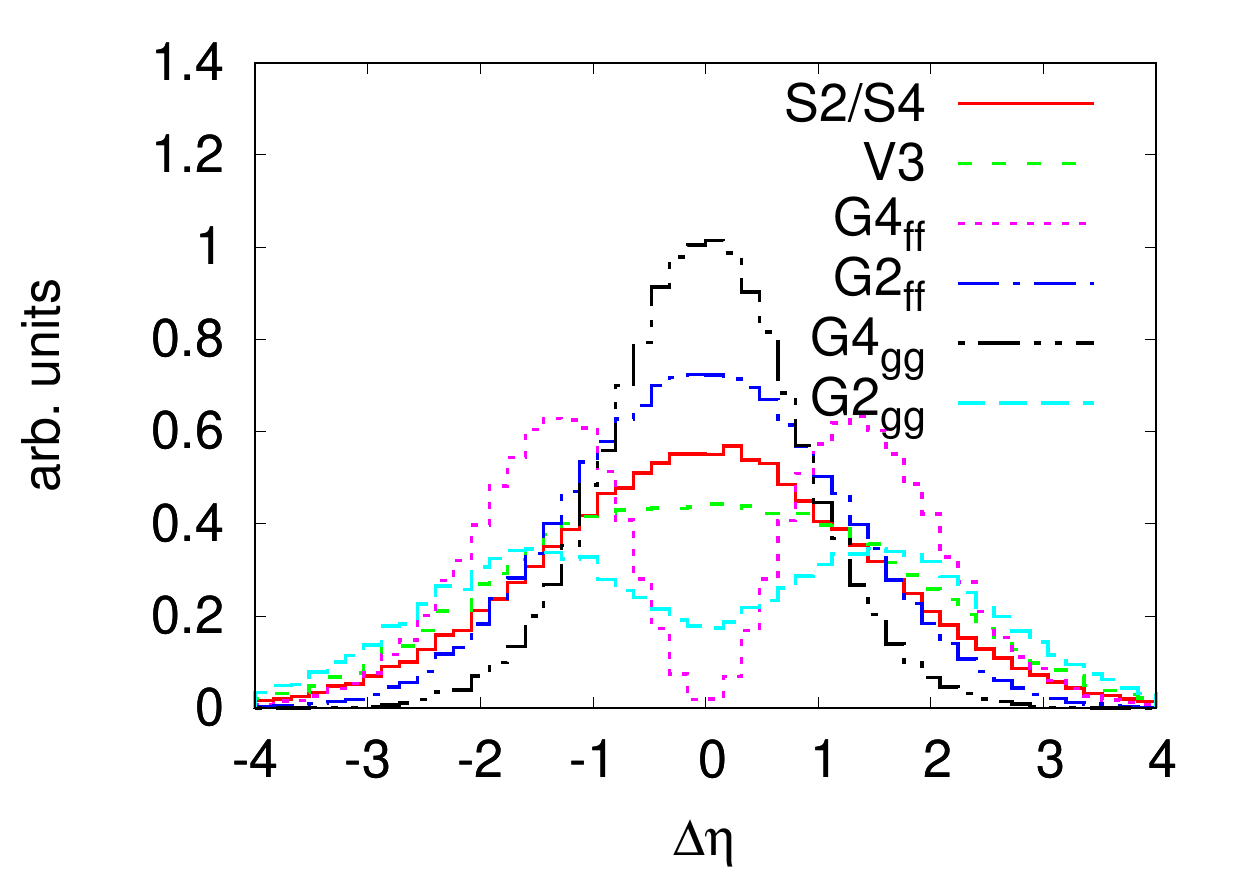} 
	\end{tabular}
	\caption{$\Delta\eta$ distribution between the two leading photon jets
          for the various models. There was very little difference between the
        S2 and S4 distributions by eye and so we have plotted them as one
        histogram.}
\label{collimated_diff}
\end{figure}

Using the $\lambda_J$ distribution of the apparent di-photon signal, we
then segregate the different scenarios into two classes: 
\begin{itemize}
	\item \textbf{Case A: a peak in signal photons at {$\lambda_J=-0.3$}}: 
Here, the distribution in Fig.~\ref{collimatedvsstandard} points
to the presence of intermediate particles $n$ and intermediate masses (of
say $m_n>15$ GeV) which lead to well resolved photons inside the photon jet,
e.g.\ V3 (40 GeV) and S4 (40 GeV) in Fig.~\ref{collimated_diff}.
There are 4 possibilities under this category: $S4,V3,G4_{gg},G4_{ff}$ (see
Table~\ref{tab:cases}). Due to the double-peak structure $V3$ can 
be distinguished from $S4,G4_{ff},G4_{gg}$ using the $\lambda_J$ distribution.
	\item \textbf{Case B: no sizeable peak at {$\lambda=-0.3$}}: Here, we
          can either have S2 or intermediate particles $n$ with a low mass. 
          Most photon pairs coming from $n$ appear as one photon
          since each from the pair hits the same ECAL cell. Thus, signal events
          resemble a conventional di-photon topology. All seven cases in
          Table~\ref{collimated_diff}
          ($S2,S4,V3,G2_{gg},G4_{gg},G2_{ff},G4_{ff}$) can lie in this
          category, depending on $m_n/M_X$.
\end{itemize}
Once the nature of the topology is confirmed by the $\lambda_J$ distribution
(i.e.\ a classification into 
case A or B), we 
then wish to determine the spin of the resonance $X$ responsible for the
excess.

\begin{figure}\begin{center}
	\begin{tabular}{c}
	\includegraphics[width=\textwidth]{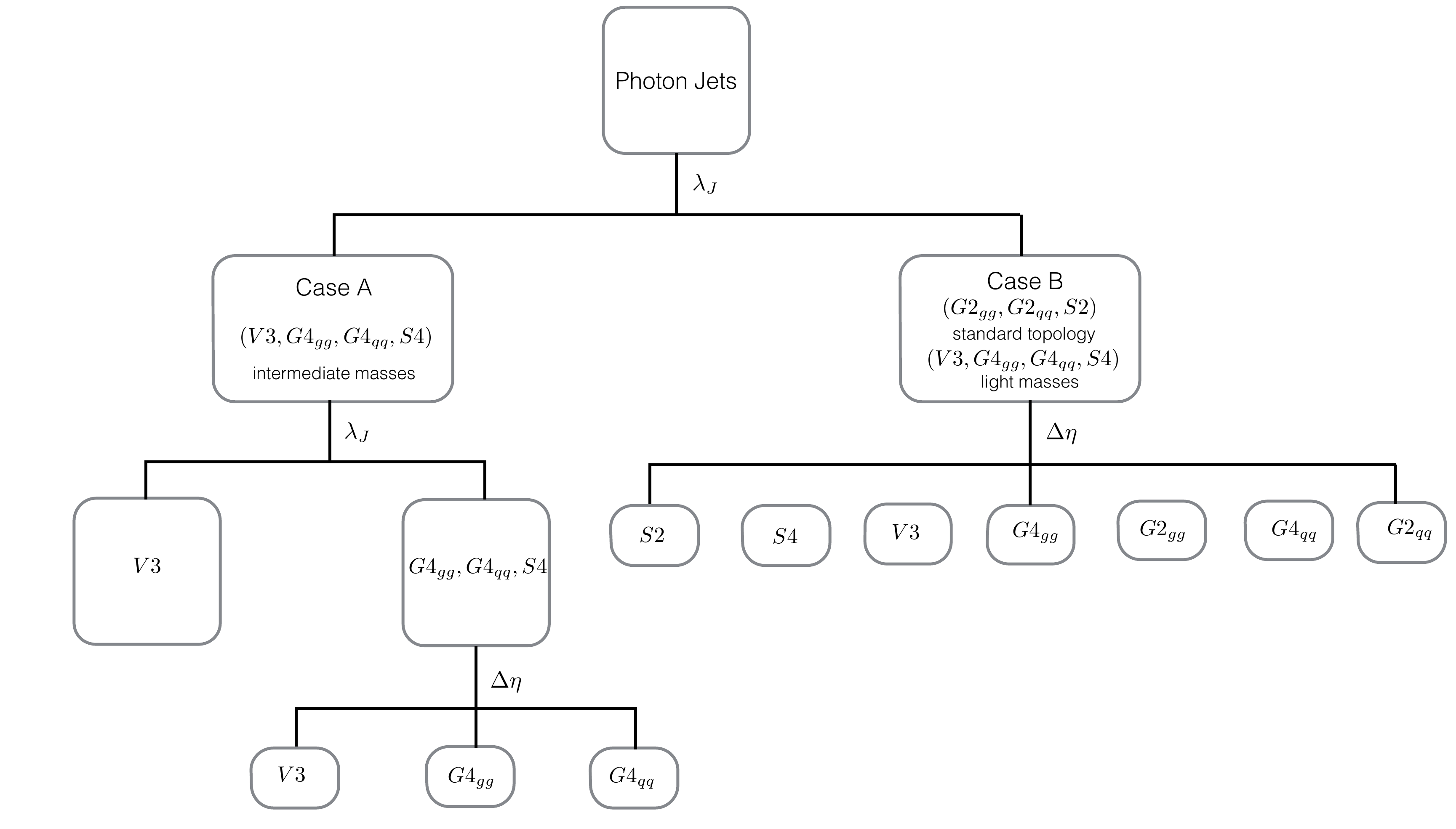}
	\end{tabular}
	\caption{Flow chart representing the analysis strategy, beginning with
          photon jets, to discern the spin of the parent resonance
          $X$.  After defining photon jets, the 
          $\lambda_J$ distribution is used to select different possibilities: 
Case A, where the $\lambda_J$ distribution
indicates the presence of intermediate $n$ particles in the decay with an
intermediate mass. Case B indicates that either the intermediate particles are
very light or absent. 
A double bump structure in the
 $\lambda_J$ distribution indicates the spin 1 ($V3$) topology. } 
\label{flowchart}\end{center}
\end{figure}

Consider case A for instance: as shown in Fig~\ref{flowchart}, the three
remaining scenarios in case A, $S4,G4_{ff},G4_{gg}$, can be distinguished from
one another by constructing the $\Delta\eta$ distribution between the leading
signal photon jets. We classify $\Delta\eta$ for a given scenario as either
central (peaking  
at zero) or non-central (two distinct peaks away from
zero) as shown in Table~\ref{tab:class}. 
We show the various distributions in Fig.~\ref{collimated_diff}. 
In the case where two scenarios can have the same $\Delta\eta$ distribution
classification (e.g.\ $S4$ 
and $G4_{gg}$), one must examine differences in the precise shapes of these
distributions to distinguish them. This will be discussed in the next
section.
\begin{table}
\begin{center}
\begin{tabular}{|c|ccccc|cc|}\hline Model & 
$S2$ & 
$S4$ & 
$V3$ & 
$G4_{gg}$ &
$G2_{ff}$ &
$G2_{gg}$ &
$G4_{ff}$ \\ \hline
$\Delta \eta$ & \multicolumn{5}{c}{Central} & \multicolumn{2}{|c|}{Non central} \\
\hline\end{tabular}
\end{center}\caption{\label{tab:class}Classification of the $\Delta \eta$
  distributions of models (listed in Table~\protect\ref{tab:cases}) as either
  central or non-central.}\end{table}
In case B, all seven models listed in
Table~\ref{collimated_diff} are possibly indicated if $m_n/M_X$ is very
small. As shown in 
Fig~\ref{flowchart}, $\Delta\eta$ will be needed to
distinguish the various models. 

\section{Spin Discrimination}
\label{kl}
The discussion in the previous section illustrates the role of
the substructure variables $\lambda_J$ and $\Delta\eta$. While
 $\lambda_J$ is useful in determining whether a given process results in
 well resolved photons in the calorimeter, $\Delta \eta$ helps
 discriminate the different spin    
hypotheses from one another.
 The signal $\Delta \eta$ distribution  changes depending upon which spins are
 involved in the chain
and they are invariant with respect to longitudinal boosts. They should
therefore be less subject to uncertainties in the parton distribution
functions (PDFs), 
which 
determine the longitudinal boost in each case\footnote{We note that whether
  the photon is in the fiducial volume or not {\em does}\/ depend upon the
  longitudinal boost, and is therefore subject to PDF errors.}. 

We wish to calculate how much luminosity we expect to need in order to be
able to discriminate the different spin possibilities in the decays, i.e.\ the
different rows of Table~\ref{tab:cases}. For this, we assume that one
particular 
Hypothesis $H_T$, is true. 
Following Ref.~\cite{Athanasiou:2006ef} (which did a continuous spin
discrimination analysis for invariant mass distributions of particle
decay chains and large $N$), we require $N$ signal events to
disfavour a 
different spin hypothesis $H_S$ to some factor $R$. We solve
\begin{equation}
\frac{1}{R} = \frac{p (H_S| N \text{~events from~}H_T)}{p(H_T | N
  \text{~events from~}H_T)} \label{ratio}
\end{equation}
for $N$, for some given $R$ (here we will require $R=20$, i.e.\ that some spin
hypothesis $H_S$ is disfavoured at 20:1 odds over another $H_T$). We are
explicitly assuming that background contributions $B$ are negligible to make our
estimate, but in practice, they could be included in the $\Delta \eta$
distributions in which case $H_S \rightarrow H_S + B$ and $H_T \rightarrow H_T
+ B$ in Eq.~\ref{ratio}.

We characterise the `$N$
events from $H_T$' by the values of a particular 
observable (or set of observables) $o_i$.
In the present paper, we shall consider the pseudrapidity
difference $\Delta \eta$ 
between the leading and next-to-leading photon jet,
$o_i^{(T)}$ (for $i \in \{ 1,2,\ldots ,N\}$) that are observed in
those events, although the observables could easily be extended to include
other observables, for example $\lambda_J$. By Bayes' Theorem, we rewrite
Eq.~\ref{ratio} as  
\begin{equation}
\frac{1}{R}=\frac{p(H_S)}{p(H_T)} \frac{p(N \text{~events~from~}H_T| H_S)}{p(N
  \text{~events from~}H_T | H_T)}  
= \frac{p(H_S)}{p(H_T)} \frac{\prod_{i=1}^N p(o^{(T)}_i | H_S)}
{\prod_{i=1}^N p(o_i^{(T)} | H_T)}. \label{oneOr}
\end{equation}
Binned data measured in the $o$
distribution $\{ n_j^{(T)} \}$ 
(for $j \in \{ 1,2,\ldots ,K\}$, $K$ being the number of 
bins), will be Poisson distributed\footnote{As argued
  above, we work in kinematic r\'{e}gimes where backgrounds can be
  neglected. We are also neglecting theoretical errors in our signal
  predictions. It 
  would be straightforward to extend our analysis to the case where some
  smearing due to theoretical uncertainties is included, where we would
  convolute Eq.~\ref{basicP} with a Gaussian distribution.}  based
on the 
expectation $\mu_j^{(X)}$ for bin $j$: 
\begin{equation} p(n_j | H_X) = \text{Pois}(n_j |
\mu_j^{(X)}), \label{basicP}\end{equation}
where $X \in \{S, T\}$ and
$\text{Pois}(n | \mu) = \frac{\mu^n e^{-\mu}}{n!}$.
Substituting this into Eq.~\ref{oneOr}, we obtain
\begin{equation}
\log\left( \frac{1}{R}\right) =
\log \left( \frac{p(H_S)}{p({H_T})} \right) + 
\sum_{j=1}^K \left[n_j^{(T)} \log \frac{\mu_j^{(S)}}{\mu_j^{(T)}} +
                                  \mu_j^{(T)} - \mu_j^{(S)}\right], \label{binned}
\end{equation}
where $\mu_j^{(T)}$ is the expectation of the number of events in bin $j$ from
$H_T$ and $n_j^{(T)}$ is a random sample of observed events obtained from
$p(n_j | H_T)$. There is a (hopefully small) amount of information lost in
going between unbinned data in Eq.~\ref{oneOr} and binned data in
Eq.~\ref{binned}. 
The first term on the right hand side contains the ratio of
prior 
probabilities of $H_T$ and $H_S$: this ratio we will set to one, having no
particular {\em a priori}\/ preference.
Then taking the expectation over many draws, $\langle n^{(T)}_j \rangle = \mu^{(T)}_j$ and
so 
\begin{equation}
\log\left( \frac{1}{R}\right) =
\sum_{i=1}^K \left[\mu ^{(T)}_j \log \frac{\mu_j^{(S)}}{\mu_j^{(T)}} +
                                  \mu_j^{(T)} - \mu_j^{(S)}\right]. \label{penultimate} 
\end{equation}
We notice that Eq.~\ref{penultimate} is not antisymmetric under $T \leftrightarrow S$,
but this is expected since we are assuming that $H_T$ is the {\em true}\/
hypothesis, in contrast to $H_S$. 
As the data come in, at some integrated luminosity, the distribution
will be sufficiently different from the prediction of some other hypothesis,
$H_S$, to discriminate against it at the level of 20 times as
 likely. 
Each term on the right-hand side is proportional to the integrated luminosity
collected $\mathcal L$,
\begin{equation}\mu_j^{(X)}=\mathcal L \sigma_{tot}^{(X)}
  \epsilon^{(X)}_j, \label{mui}
 \end{equation}
where $\sigma_{tot}^{(X)}$ is the assumed total signal cross-section (i.e.\
the $X$ production cross section) before
cuts for $H_X$
and $\epsilon_j^{(X)}$ is the probability that a signal event makes it past all
of the cuts and into bin $j$, under hypothesis $X$. Assuming that
$\sigma_{tot}^S=\sigma_{tot}^T \equiv \sigma_{tot}$,
we may solve Eq.~\ref{penultimate} and Eq.~\ref{mui} for ${
  N}_R=\mathcal L \sigma_{tot}$, the expected number of 
total signal events  
required to disfavour $H_S$ over $H_T$ to an
odds factor of $R$:
\begin{equation}
N_R = \frac{\log  R}{\sum_{j=1}^K
  \left[\epsilon ^{(T)}_j \log \frac{\epsilon_j^{(T)}}{\epsilon_j^{(S)}} + 
                                  \epsilon_j^{(S)} - \epsilon_j^{(T)}\right]}.
\label{lr}
\end{equation}
One property of this equation is that if $\epsilon^{(T)}_j=\epsilon^{(S)}_j$
$\forall$ $j$, then ${\mathcal L}_R \rightarrow \infty$. This makes sense:
there is no luminosity large enough such that it can discriminate between
 identical distributions.  
Eq.~\ref{lr} works for multi-dimensional cases of several observables: 
one simply gets more bins for the multi-dimensional case. 
If one works in the large statistics limit, for continuous data (rather than
binned data), one obtains a required number of events that is
related~\cite{Athanasiou:2006ef} to the 
Kullback-Leibler divergence instead~\cite{kullback1951}. 
The Kullback-Leibler divergence is commonly used when one has analytic
expressions for distributions of the observables (see
Ref.~\cite{Athanasiou:2006ef}), and has 
the advantage of utilising the full information in $o$. 
We do not have analytic expressions, partly because they depend upon parton
distribution functions, which are numerically calculated. 
Our method loses some information by binning, but it has the considerable
advantage that it includes kinematical selection and detector effects (all
contained within the $\epsilon_j$). Eq.~\ref{lr} has the property that: if one
halves the total $X$ production cross-section, one requires double the
luminosity to keep the discrimination power (measured by $R$) constant. 

Since we shall
estimate $\epsilon^{(X)}_j$ numerically via Monte-Carlo event generation,
there is a potential problem we have to deal with: a bin might end 
up 
with no generated events and so one encounters divergences from the logarithm
in the 
denominator of Eq.~\ref{lr}. This is due, however, to not using enough Monte
Carlo statistics, where $M$ signal events are simulated in total for each
parameter choice and for each hypothesis pairing. 
We restrict the range of $o$ and use large enough Monte Carlo statistics
($M=200000$) such that no bins (that are set to be wide enough) contain
zero events. 

\subsection{Event Selection and Results}
Using the statistic developed in Eq.~\ref{lr}, we first first discriminate
Case A from B defined in 
Section~\ref{variables}. Thus, in the event of an apparent di-photon excess in a certain invariant mass bin say $m^{(0)}_{\gamma\gamma}$, we propose the following steps:
\begin{itemize}
	\item We relax the isolation criteria and re-analyse the events by constructing photon jets.
	\item The invariant mass $m_{j_1 j_2}$ of the two leading photon jets
          for each 
          events are required to lie around $m^{(0)}_{\gamma\gamma}$:
 we require  $1100<m_{j_1 j_2}/\textrm{GeV}<1300$. 
	\item Photon jets from pions are eliminated by requiring
          that leading jet to have no 
          tracks ($n_T=0$) and by requiring
          $\log \theta_J<-0.5$. We also take into account the photon
          conversion factor. 
	This depends on whether the photon  converts before or after exiting
        the pixel detector. This conversion probability is a function of the
        number of radiation lengths ($a$) a photon passes through before it
        escapes  the first pixel detector and is given by~\cite{Ellis:2012zp}
	\begin{equation}
	P(\eta)=1-\exp(-\frac{7}{9}a(\eta)).
	\end{equation}
	We approximate this by an $\eta$ independent conversion probability
        $P(\eta)=0.2$. 
	\item The substructure of each jet is analysed using $\lambda_J$ to
          determine whether it is in Case A or B. 
\end{itemize}
Fig.\ref{flowchart} gives a pictorial representation of these steps. 
We use $m_n=40$ GeV and $m_n=1$ GeV as examples for the model hypotheses to be
tested. 
We simulate $2\times 10^5$  events for the topologies predicted by $H_T$
and $H_S$  and compute $\lambda_J$ for the all events
which pass the basic selection criteria. 
To avoid any zero event bins, $\lambda_J$ is binned between
$[-4,0]$ with a bin size of $0.6$ and the efficiency for each
particular bin is extracted for both distributions from the simulation. 
Owing to the distinct nature of the $\lambda_J$ distribution for both the
cases, 3-4 events is sufficient to discriminate between case A and case B. 
The $m_n=1,40$ GeV cases both have a post-cut acceptance efficiency of $\sim
55\%$.  
For a cross-section of 0.5 fb, we can accumulate 
some five signal events with $\sim$18 fb$^{-1}$ of integrated luminosity.  
Once the nature of the topology (corresponding to a given case) is identified,
our next step is to discriminate the different possibilities within it. 
Both of the scenarios are handled independently as follows:\\ 
  \textbf{CASE A}: In this case there are only four possibilities
  corresponding to a multi-photon topology (i.e.\ proceeding through an
  intermediate $n$). As discussed earlier, we do not impose the requirement of
  two 
  isolated photons, since the photons from $n$ tend to fail isolation cuts. We
  compute $\Delta\eta$ between the two leading photon jets. 
In order to discriminate V3 from the other cases, the twin-peaked structure of
$V3$ under 
 $\lambda_J$ (as shown in Fig~\ref{lambdaj}) can be employed to discriminate
 it collectively 
 from $S4,G4_{gg},G4_{ff}$. In this case one requires a minimum of 20 signal
 events to disfavour the other three at a $20:1$ odds. 
 All samples are characterised by a minimum of $\sim 55\%$ acceptance
 efficiency. With this information then, one can disfavour
 $S4,G4_{gg},G4_{ff}$ in favour of $V3$ with  $\sim$ 72~fb$^{-1}$ of integrated
 luminosity for a
 0.5 fb signal cross-section.  
 
 $S4,G4_{gg},G4_{ff}$ can then be discriminated from one another using
 $\Delta\eta$ between the two leading jets. Table~\ref{tab:Spindiscrimination1lambda}
 computes the minimum number events required for pairwise discrimination of
 the three cases for $m_n=40$ GeV and is computed using Eq.~\ref{lr} 
 To avoid zero event bins in the $\Delta \eta$ distribution, we 
 restrict the {\em a priori}\/ range of $|\Delta \eta| \in [-5,5]$ to
 $[-4,4]$. As shown in the Table~\ref{tab:Spindiscrimination1lambda}, disfavouring
 $S4$ as compared to $G4_{gg}$ constitutes the largest expected number of
 required signal events 
 \textit{i.e.} 29. This can be achieved with a luminosity of
 $\sim$ 105~fb$^{-1}$. Thus in the event of a discovery corresponding to Case A, it
 is possible to get exact nature of the spin of $X$ within 105 fb$^{-1}$ of data.
 \begin{table}[h!]
 	\begin{center}
 		\begin{tabular}{|c|c|c|c|} 
 			\hline
 			
 			$N_R$           & $S4$   & $G4_{gg}$ &  $G4_{ff}$ \\ \hline
 			\hline
 			
 			$S4$          &$\infty$&22&13\\ \hline

 			$G4_{gg}$     &29&$\infty$&4\\ \hline  
 			
 			$G4_{ff}$     &19&5&$\infty$ \\ \hline    
 		\end{tabular}
 		\caption{\label{tab:Spindiscrimination1lambda} Spin discrimination: $N_R = {\mathcal L}
 		\sigma_{tot}^{(X)}$, the expected
 		number of total signal events required to be produced 
 		to discriminate against the `true' row
 		model versus a column model by a factor of 20 at the 13
                TeV 
 		LHC for $m_n=40$ GeV.} 
 	\end{center}
 	\end{table}
 
\textbf{CASE B:} This constitutes the more complicated of the two cases. 
Since the two hard photons inside the photon-jet for the multi-photon topologies
can not be well resolved, the substructure is similar to the conventional
single photon jet from the standard di-photon topology. Thus there are more
cases to distinguished in this case. 
We compute the $\Delta\eta$ between the leading two jets of the event. To
avoid zero event bins in the $\Delta \eta$ distribution, we  
restrict the {\em a priori}\/ range of $\Delta \eta$ from $[-5,5]$ to
$[-4,4]$. 

The signal models here are characterised by an acceptance efficiency of at least
$55\%$. Using the cross-section of 0.5 fb, 
we find that the cases $S2$ and $S4$ are virtually indistinguishable owing to
the similar shapes of their $\Delta\eta$ distributions. They thus cannot be
distinguished on the basis of the $\Delta \eta$ distribution. 
However, as shown in Fig.~\ref{collimatedvsstandard}, the presence of
secondary bump for the collimated case will help in distinguishing these two
cases. In this case, the same technology we have developed for the $\Delta
\eta$ distribution could be employed for the $\lambda_J$ distribution.
 
 Distinguishing $S2,S4$ from $V3$ requires a maximum expected number of
 events of 250-300.  
  This is achievable with 1.1 ab$^{-1}$ of integrated luminosity, assuming an
  acceptance of $\sim 55~\%$ and a signal production cross-section of 0.5
  fb. Distinguishing  scenarios like $S2$ from $G4_{ff}$ or $G4_{gg}$
  requires 23 events or less: these could be discriminated with
  $\sim$84~fb$^{-1}$
  for our reference cross-section of 0.5 fb, whereas
the rest of the pairs of spin hypotheses can be distinguished within 364~fb$^{-1}$ of data.  
  
\begin{table}[h!]
	\begin{center}
		\begin{tabular}{|c||c|c|c|c|c|c|c|} 
			\hline
                  $N_R$       & $S2$    & $S4$ & $V3$ & $G2_{gg}$ & $G4_{gg}$ & $G2_{ff}$ & $G4_{ff}$ \\ \hline
		\hline
			$S2$       & $\infty$&$>2000$     &272&27&15&91&14  \\ \hline 
			$S4$       &  $>2000$   &$\infty$ &255&26&15&96&13\\ \hline   
			$V3$       & 260        &248&$\infty$ &54&9&37&21 \\ \hline 
			$G2_{gg}$  &  32    &31&65&$\infty$ &5&13&38      \\ \hline 
			$G4_{gg}$  &   23   &24&14&6&$\infty$ &54&4\\ \hline  
			$G2_{ff}$  &   102   &110&44&12&40&$\infty$ & 8   \\ \hline 
			$G4_{ff}$  &   19   &18&28&37&5&12&$\infty$ \\ \hline    
		\end{tabular}
		\label{Spindiscrimination}
	\end{center}
	\caption{Spin discrimination of two models: $N_R = {\mathcal L}
          \sigma_{tot}^{(X)}$, the expected
          number of total signal events required to be produced 
          to discriminate against the `true' row
          model versus a column model by a factor of 20 at the 13 TeV
          LHC for $m_n=1$ GeV.} 
       \end{table}
 
\section{Mass of the intermediate scalar}
\label{mas}
\begin{figure}[h]
	\begin{center}
          \includegraphics[width=
          0.8 \textwidth]{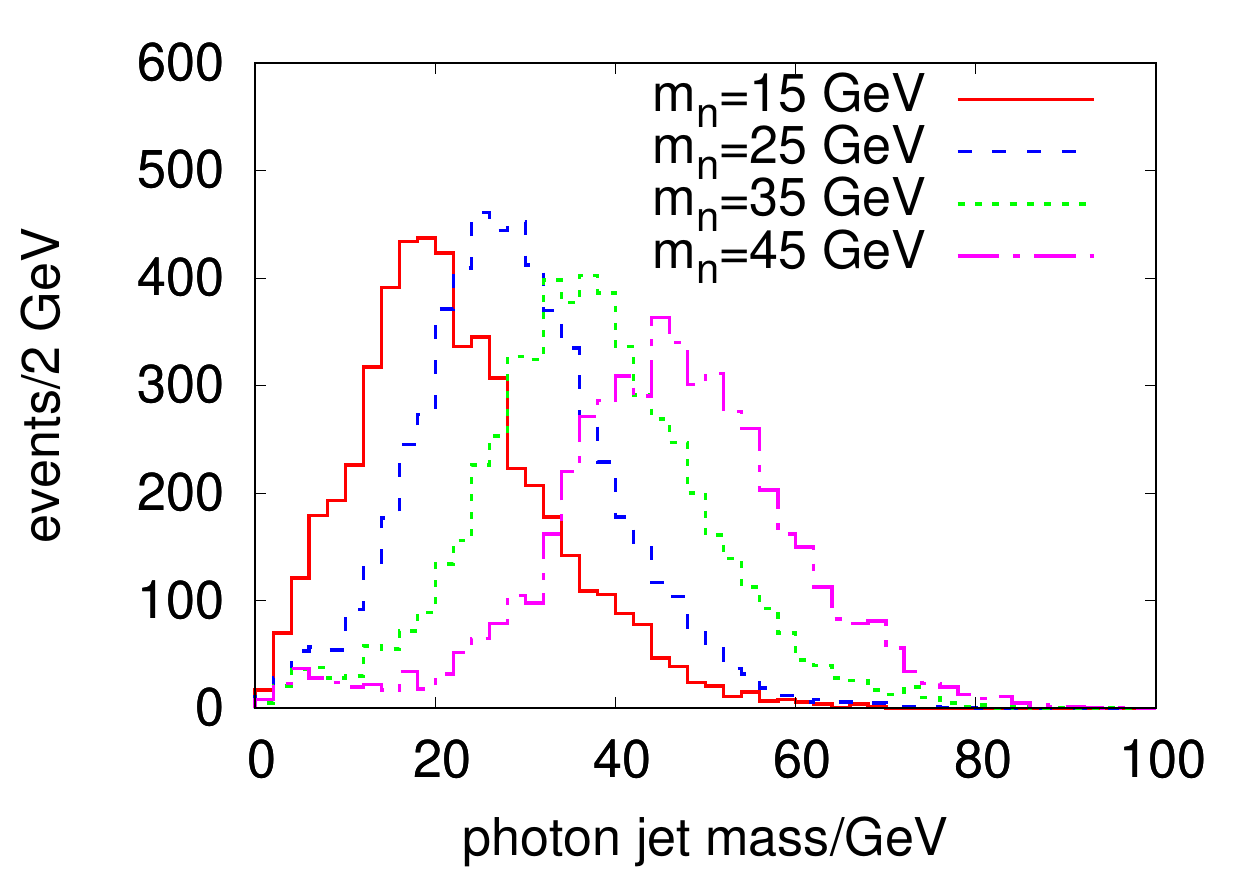}
          \caption{Comparison of the S4 photon jet mass
            distributions for 
            the leading photon jets and various $m_n$.} 
          \label{jetmass}
	\end{center}
\end{figure}

A multi-photon topology is indicative of the presence of two scales in the
theory: $m_X$ and $m_n$.
While the scale of the heavier resonance is evident from the apparent
di-photon invariant mass distribution, extracting the mass of the
lighter state may be more difficult. 
From Fig.~\ref{comparison}, we see that
for low to intermediate masses, one does not obtain
isolated photons from $n$ which may be used to reconstruct its mass. We
therefore examine the invariant mass of photon
jets. 
The decay constituents of $n$ retain its properties such as its $p_T$,
pseudo-rapidity $\eta$, mass 
\textit{etc.}. Fig.~\ref{jetmass} shows a comparison of the mass of the leading
jet for $S4$ and a few different values of $m_n$.
The peak of each distribution, which can be fitted, clearly tracks with the
mass of $n$. Using an estimate based on the statistical measure introduced in
section~\ref{kl}, we calculate that 25 signal events would be required to
discriminate the 35 GeV from the 45 GeV hypothesis, for instance: i.e.\ $\sim$91
fb$^{-1}$ 
of integrated luminosity and a signal cross-section of 0.5 fb. 
Thus, for intermediate masses and reasonable amounts of integrated luminosity,
a fit to the peak should usefully constrain $m_n$, at least for $m_n\gsim 10$ GeV.

\section{Conclusion}
\label{conclusion}
In the event of the discovery of a resonance at high di-photon invariant
masses, it will of course be important to dissect it and discover as much
information about its anatomy as possible.  
Here, we have provided a use case
for Refs.~\cite{Ellis:2012zp,Ellis:2012sd}, where photon jets, photon sub-jets
and simple kinematic 
variables were defined that might provide this information. The apparent
di-photon signals may in 
fact be multi-photon (i.\.e greater than two photons), where several photons
are collinear, as is expected when intermediate particles have a mass much
less than the mass of the original resonance. 
We identified useful variables for this purpose: the pseudorapidity difference
between the photon jets helps discriminate different spin combinations of the
two new particles in the decays. 
We quantify an estimate for how many signal events are expected to be required
to provide discrimination between different spin hypotheses, setting up a
discrete version of the Kullback-Leibler divergence for the purpose. For the
discovery of a 1200 GeV resonance with a signal cross-section of 0.5 fb, many
of the spin possibilities can be discriminated within the expected total
integrated luminosity expected to be obtained from the LHC.
A simple sub-jet variable $\lambda_J$ provides
a good discriminant between the di-photon and multi-photon cases. 
The invariant mass of the individual photon jets provides useful information
about the intermediate resonance mass.

We hope that our study motivates work from the experimental
collaborations, that have access to detailed detector information. For 
example, it would be interesting to see how much `layer 1' of ATLAS' ECAL
would help verify the very light $n$ cases. Also, photon conversion rates
would be different for two almost collinear photons than for a single photon,
providing another possible tool for diagnosing multi-photon final states.

\section*{Acknowledgements}
This work has been partially supported by STFC ST/L000385/1.
We thank the Cambridge SUSY Working group, Sandhya Jain and Kerstin Tackmann
for helpful 
comments and discussions.
BA and AI  would like to thank the organisers of Rencontres de Moriond
2016  where the project was conceived. AI would also like to thank the
hospitality of The University of Cambridge and King's College London where
different 
aspects of the project were discussed. We would also like to thank the
Aspen Center for Physics and the
organisers of `From Strings to LHC'- IV where parts of the project were
discussed.
\appendix
\section{Signal Model parameters \label{sigmod}}
We now detail the model parameters picked for
each case for our numerical simulations. 
Firstly we specify the
$X\rightarrow \gamma \gamma$ case, where we choose
$\eta^{-1}_{GX}=40$ TeV and $\eta^{-1}_{\gamma X}=80$ TeV.
When $X$ is spin 2 and we consider fermion anti-fermion production,
$\eta^{-1}_{T\psi X}=\eta^{-1}_{T \gamma X}=40$ TeV.
When  $X$ is spin 2 and we consider gluon gluon production,
$\eta^{-1}_{T G X}=\eta^{-1}_{T \gamma X}=80$ TeV.

When instead, we consider intermediate scalar $n$ particles in the decays of
$X$, we fix $\eta_{n \gamma \gamma}^{-1}=\eta_{GX}^{-1}=20$ TeV for the spin 0
$X$ case. For spin 1 $X$, $\eta_{n X \gamma}=0.3/(10 \textrm{~TeV})$ and
$\eta_{n \gamma \gamma}^{-1}=10$ TeV. For spin 2 $X$ and fermion anti-fermion
production, $\eta^{-1}_{Tn X}=\eta_{n \gamma \gamma}^{-1}=10$ TeV and
$\eta^{-1}_{T \psi X}=20$ TeV. For spin 2 $X$ and glue glue production, 
$\eta^{-1}_{Tn X}=\eta_{n \gamma \gamma}^{-1}=20$ TeV and $\eta^{-1}_{TG
  X}=40$ TeV.
\bibliography{spin.bib}
\end{document}